\documentclass[pdflatex,sn-basic]{sn-jnl}
\usepackage{etoolbox}
\usepackage{algorithm,algorithmic}
\usepackage{amsmath,amssymb}
\usepackage{booktabs,array,tabularx}
\usepackage{graphicx}
\usepackage{adjustbox}
\makeatletter
\newenvironment{widetable}[1][]{\begin{tableorg}[#1]\centering\tablebodyfont
  \renewcommand\footnotetext[2][]{{\raggedright\removelastskip\vskip3pt\let\tablebodyfont\tablefootnotefont
  \hskip0pt\if!##1!\else{\smash{$^{##1}$}}\fi##2\par}}}{\end{tableorg}}
\makeatother
\usepackage{tikz}
\usetikzlibrary{arrows.meta}
\usepackage{placeins}
\usepackage{xcolor}
\usepackage{xurl}
\newcolumntype{L}[1]{>{\raggedright\arraybackslash}p{#1}}
\hypersetup{hidelinks,pdftitle={What Should an LLM Know? Knowledge-Guided Warm Starts for Software Optimization},pdfauthor={Srinath Srinivasan and Tim Menzies}}

 \usepackage[most]{tcolorbox}
\newtcolorbox{rqans}{enhanced, breakable, frame hidden, colback=white,
  borderline west={1.2pt}{0pt}{gray!60}, sharp corners,
  left=8pt, right=0pt, top=1pt, bottom=1pt,
  enlarge left by=1.2em, width=\linewidth-1.2em,
  before skip=2pt, after skip=8pt}
\begin{document}
\apptocmd{\thebibliography}{\raggedright}{}{}

\title[What Should an LLM Know?]{What Should an LLM Know? Knowledge-Guided Warm Starts for Software Optimization}
\author[1]{\fnm{Srinath} \sur{Srinivasan}}\email{ssrini27@ncsu.edu}
\author*[1]{\fnm{Tim} \sur{Menzies}}\email{timm@ieee.org}
\affil[1]{\orgname{North Carolina State University},\ \orgaddress{\city{Raleigh},\ \state{North Carolina},\ \country{USA}}}

\abstract{ 
    Large language models (LLMs) process gigabytes in minutes; human experts work far
    more slowly. Prior studies that explored combinations of human/AI optimizers ignore this gap. Those studies
    asked systems or humans to label
    too many examples (e.g., SynthCore needs over four dozen per trial). Here, we
    test human and AI knowledge under a budget where evaluation is
    practical (a few dozen labels or fewer).
    
    We compare five baselines (random search, RRP, EZR, few-shot prompting, and a budget-matched SynthCore variant) against three ways to give an LLM domain knowledge before it proposes its
    first configurations: (1)~AMP lists the domain's rules, then checks each
    candidate against them; (2)~HKMA adds scouted data or documentation to the
    prompt; and (3)~H-DKP has human experts correct the LLM's rules.
    
    We find that domain knowledge improves the ``start-up'' phase of LLM optimization, but only
    when that knowledge is checked, by the LLM itself or by a human expert. Across 49
    datasets, unchecked knowledge did no better than few-shot prompting or EZR.
    On the other hand, AMP's self-checking variant (AMP-4) reached the best statistical rank in 44 of
    49 datasets (budget-matched SynthCore: 23 of 49), and H-DKP did so in 7 of
    8 expert studies.
    These checks cost little: half a cent of LLM calls per trial, or 6 to 10 expert
    contacts. Hence we say LLMs may not need more data, but rather their knowledge
    needs checking (and checking is cheap).}

\maketitle

\section{Introduction}\label{sec:introduction}
In this paper we assess the value of   teaming up humans with LLMs to configure software.
Modern software systems expose numerous tunable parameters that govern
competing objectives: a database configuration trades query latency against
memory footprint, while a machine learning pipeline balances predictive
accuracy against training time \citep{harman2012search,VanAken2017,fu2016tuning}. Configuration is a complex task: configuration errors are widespread, and
parameter spaces continue to grow in complexity
\citep{zhou2011understanding,xu2015hey}. Accordingly we ask:
\begin{center}
{\em How to better control and configure complex software?}
\end{center}

One approach is to configure systems using the knowledge contained inside large language models (LLMs). 
For example, in the ``warm start''
approach explored by~\citep{senthilkumar2026synthcore}, some
oracle1 (human or AI) passes a few labeled examples to a second oracle2. In this paper, a \emph{label} is the measured objective scores of one
configuration (e.g., the runtime and memory use of one database setting);
we also call this an \emph{evaluation}. Labels are costly since each
one needs the system to be run, or an expert to be asked.

Repeated studies show that if oracle1 can generate good labels, oracle2's final performance can be dramatically improved; e.g., see the SynthCore study and prior work~\citep{senthilkumar2026synthcore,6693104}.

This paper addresses two major drawbacks with the SynthCore study:
\begin{itemize}
\item That prior analysis was fully automatic. Here, we conduct studies that combine AI and
human decision making~\citep{ramirez2019interaction,takagi2001interactive}\footnote{Human experimentation was approved by the NC State University Institutional Review Board (protocol \#28893); see Section~\ref{sec:human-study} for participant, consent, and protocol details.}.
\item
That prior analysis made unreasonable demands on the amount of available labeled data. SynthCore
reduced conclusion variance by ensembling 15
prompting sessions, but it needs dozens of distinct labels per trial (on average). Here, we make conclusions using just 24 labels.
\end{itemize}

% \color{red}

% (Note to reviewers, to be removed before publication: This paper reports the results of the study registered at MSR’26~\citep{registeredreport2026}. Compared to that report, this paper adds (a) much more background on methods and motivation and (b) the promised results. We submit it here as a regular paper because other commitments meant we missed the deadline for the special issue on MSR’26 registered reports.)

% Note that DAPR was excluded from this study due to its label-intensive nature.
% We could not find an appropriate surrogate for DAPR in the low-budget setting studied here.
% \color{black}

There are many reasons why it is prudent to study configuration using very limited budgets: 
\begin{itemize}
\item {\em Experts are slow} at labeling~\citep{menzies1992expert,easterby1980design,KingtonAlison2009,lustosa2024learning,valerdi2010heuristics}.
  Knowledge engineers in the 1980s expected only 5--10 rules per expert
  per day \citep{menzies1992expert}. And if rushed, the labels offered by experts become inaccurate~\citep{easterby1980design}. Interactive optimizers face the same limit: users tire after only a few judgments~\citep{takagi2001interactive} (see also Section~\ref{sec:hitl}).
\item {\em Automatic labeling methods are heuristic and error prone}. Regular-expression heuristics are
  crude \citep{kamei2012large}; quality labeling tools like SZZ have a long list of known problems~\citep{herbold2022problems}, and LLMs can assist but not replace human
  labelers \citep{ahmed2025llmsreplacemanualannotation}.
\item {\em Execution-based oracles can be slow}: some labels need a compile and a
  benchmark run of
  slow and complex software, such as a full test suite \citep{nair2017using,chen2018sway}.
\item {\em Evaluation is costly}: exploring the 11 parameters of (e.g.) the x264 video encoder takes
  1,536 hours \citep{valov2017transferring}, so budgets are often limited
  to dozens of evaluations \citep{nair2020finding,chen2021efficient}.
\item {\em Configuration spaces are vast}: MySQL's 460 binary options define $2^{460}$
  configurations \citep{xu2015hey}. Even far larger budgets sample only a tiny part of such spaces.
\item {\em Cheap labels are often wrong}: historical logs can offer many misleading labels.
  For example, 90\% of technical-debt ``false positives'' can be
  incorrect \citep{yu2020identifying}, with similar issues in security
  \citep{wu2021data}, static analysis \citep{kang2022detecting}, and
  defect data \citep{shepperd2013data}.
\end{itemize}
For these reasons, this paper focuses on very small budgets: 24 labels.
At this scale, the human effort is realistic: it can fit within a standard work week without disrupting it\footnote{On our expert study (Section~\ref{sec:human-study}),
each specialist reviewed and refined the LLM's domain rules across 6--10 short email exchanges over 5--17 calendar days (up to 19 lines for the initial claim review, and 1--3 lines per follow-up reply).}. This focus sets our work apart from much prior configuration and
hyperparameter optimization research, which assumes (a)~no human experimentation and (b)~hundreds to thousands
of evaluations
\citep{NIPS2011_86e8f7ab,bergstra2012random,VanAken2017,pmlr-v188-pfisterer22a}.

\begin{figure}[t]
\centering
\resizebox{\linewidth}{!}{%
\begin{tikzpicture}[font=\footnotesize\sffamily,
  box/.style={draw=gray!70, rounded corners=2pt, align=center, inner sep=5pt, minimum height=1.9cm}]
\node[box, fill=blue!6,  text width=2.4cm] (in)   at (0,0)    {\textbf{Inputs}\\[2pt] task description\\ 4 labeled examples};
\node[box, fill=green!7, text width=4.6cm] (know) at (4.55,0) {\textbf{Add domain knowledge} (one of)\\[2pt]
  AMP: list rules, check each candidate\\ HKMA: scouted data or documents\\ H-DKP: expert-reviewed rules};
\node[box, fill=orange!8, text width=2.6cm] (warm) at (9.35,0) {\textbf{LLM warm start}\\[2pt] proposes 4\\ configurations\\ (2 better, 2 poorer),\\ then label them};
\node[box, fill=orange!8, text width=2.6cm] (smo)  at (9.35,-2.75) {\textbf{Sequential optimizer}\\[2pt] selects and labels\\ 16 more configurations\\ (24 labels in total)};
\node[box, fill=gray!8,  text width=6.4cm] (eval) at (3.4,-2.75) {\textbf{Evaluate}\\[2pt] quality: Chebyshev distance to the ideal point,\\ ranked by Scott--Knott\\[2pt] cost: labels, LLM calls, expert emails};
\draw[-latex, thick, gray!70!black] (in) -- (know);
\draw[-latex, thick, gray!70!black] (know) -- (warm);
\draw[-latex, thick, gray!70!black] (warm) -- (smo);
\draw[-latex, thick, gray!70!black] (smo) -- (eval);
\end{tikzpicture}}
\caption{Our trails starts with 4 labeled examples. Domain knowledge may be added  before an LLM proposes 4 initial configurations. A sequential optimizer then selects the rest, for 24 labels in total.}
\label{fig:warm-start-overview}
\end{figure}
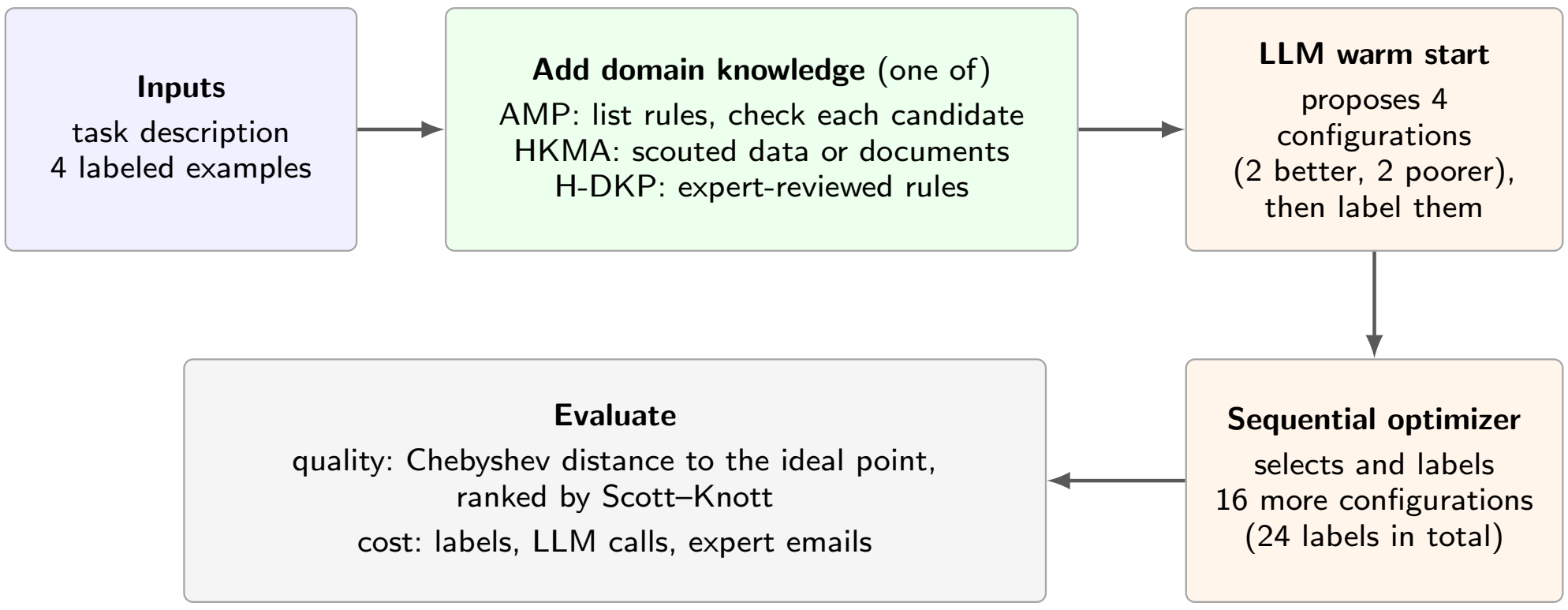

Within this budget, we compare three methods for giving an LLM domain knowledge before it proposes its first configurations (Figure~\ref{fig:warm-start-overview}). AMP has the LLM list the domain's rules, then check each candidate against them~\citep{khot2022decomposed,madaan2023selfrefine}. HKMA adds scouted data or documentation to the prompt~\citep{NIPS2011_86e8f7ab,lewis2020retrieval}. H-DKP has human experts correct the LLM's rules~\citep{schreiber2000kads}. Together, these methods draw on four sources of domain knowledge: the LLM itself, scouted data, external documents, and human experts.

We test these methods on 49 SE multi-objective optimization
problems~\citep{mootrepo}. We ask:

{\bf RQ1:} With only 24 labels, can domain knowledge help an LLM propose
better initial configurations?
\begin{rqans}
\textbf{Only if that knowledge is checked.} Unchecked knowledge
(LLM-discovered rules, scouting, documents) did no better than direct
few-shot prompting or the EZR active learner. But AMP-4, which checks each
candidate against the rules, reached the best statistical rank on 44 of 49
datasets.
\end{rqans}\noindent
{\bf RQ2:} Does increasing the number of configuration parameters change
which kinds of knowledge help?
\begin{rqans}
\textbf{Yes.} Checking matters most in high dimensions, where AMP-4 ranked
best on 16 of 17 tasks, while documentation-only prompting fell from 12 of
15 (low) to 9 of 17 (high).
\end{rqans}\noindent
{\bf RQ3:} Does human review of the LLM's domain rules help?
\begin{rqans}
\textbf{Yes.} H-DKP ranked best on 7 of 8 expert tasks, and beat direct
prompting on 7 of 8. However, EZR also ranked best on 7 of 8, at zero cost.
\end{rqans}
\noindent
{\bf RQ4:} What does checked knowledge cost?
\begin{rqans}
\textbf{Very little.} AMP-4 costs half a cent of LLM calls per trial ($\le 24$ labels) and H-DKP needs only 6 to 10 short expert emails. By contrast, restricting multi-session SynthCore to a comparable budget (SynthCore-28, 26.31 labels) drops its win rate to 23 of 49 tasks, trailing both AMP-4 (39 of 49) and zero-cost EZR (32 of 49).
\end{rqans}
\noindent
In summary, this paper makes five contributions:
\begin{itemize}
\item Evidence that domain knowledge improves LLM warm starts only when it
  is checked (Sections~\ref{sec:rq1} and~\ref{sec:rq2}).
\item A human-in-the-loop study on eight tasks across nine expert \emph{tracks} (Sections~\ref{sec:human-study} and~\ref{sec:rq3}).
\item A cost analysis across configuration labels, LLM calls, and expert
  effort (Section~\ref{sec:rq4}).
\item A methodological lesson: LLM methods must be compared against simple
  baselines, since the model-free EZR active learner often does as well
  (Sections~\ref{sec:rq3} and~\ref{sec:discussion}).
\item A replication package containing all our code, data, and prompts at \url{https://github.com/srina1h/What-Should-an-LLM-Know}.
\end{itemize}

\section{Background and related work}\label{sec:background}

\subsection{Configuration search with few observations}
Software configuration strongly affects performance across database management, software analytics, and compiler tuning. \citet{VanAken2017} use runtime measurements to recommend database settings, while studies in software analytics show that learner hyperparameters substantially alter prediction quality \citep{fu2016tuning,Tantithamthavorn16}. Although simple heuristics sometimes suffice \citep{agrawal2019dodge} and coarse performance models can still rank configurations effectively \citep{nair2017using}, performance models often transfer poorly across hardware platforms and workloads \citep{valov2017transferring}. Consequently, optimizers must evaluate candidate configurations directly on the target system.

Because physical evaluations are costly, sequential optimizers select new configurations iteratively from a small set of prior observations. FLASH applies sequential model-based optimization to software configuration \citep{nair2020finding}, and Bayesian optimization is widely used for compiler autotuning \citep{chen2021efficient}. Tree-structured Parzen Estimation (TPE) finds promising and unpromising regions separately to guide candidate selection \citep{NIPS2011_86e8f7ab}. Other works study sampling-based baselines \citep{chen2018sway}, landscape analysis from minimal data \citep{lustosa2024learning}, and lightweight active learning via the EZR toolkit \citep{menzies2026can}. These studies define the data-light software engineering challenge \citep{ganguly2025low}: find near-optimal configurations within about two dozen evaluations.

\subsection{LLMs for optimization}
Recent work explores using LLMs to guide black-box optimization. Several frameworks invoke the language model repeatedly throughout the search loop: LLAMBO uses LLMs for warm starts, surrogate modeling, and candidate sampling in Bayesian optimization \citep{liu2024llambo}, LMEA employs an LLM as a crossover and mutation operator in evolutionary search \citep{liu2024lmea}, and \citet{huang2024exploring} evaluate iterative LLM optimization across continuous and discrete benchmarks. While iterative LLM-in-the-loop search can be effective, invoking a large model at every optimization step increases inference latency and token costs.

An alternative is to invoke the LLM only at initialization to warm-start a fast sequential optimizer. \citet{senthilkumar2024largelanguagemodelsimprove} show that prompting an LLM with task metadata and a few labeled examples can produce useful initial configurations; their direct few-shot approach serves as our BS-LLM baseline. However, single-pass few-shot generation is often unreliable when parameter interactions are complex. To improve robustness, SynthCore ensembles candidates over 15 independent few-shot generation sessions \citep{senthilkumar2026synthcore}. Although ensembling reduces variance, it needs more than twice our 24-label budget. Despite growing interest in LLM-assisted optimization \citep{liu2024lmea,huang2024exploring}, it remains unclear which knowledge sources improve optimization quality when labels are scarce, whether passive context injection is sufficient or active constraint verification is required, and when the additional inference tokens or expert review effort are justified over lightweight model-free optimizers. Rather than scaling up the labeling budget through multi-session ensembling, we investigate whether structuring and verifying domain knowledge allows an LLM to generate stronger warm starts within a strict 24-label budget.

\subsection{Sources of domain knowledge}
Knowledge engineering examines how domain information is elicited, structured, and verified for computational use \citep{schreiber2000kads}. In software configuration, relevant domain knowledge encompasses feature importance, feasible parameter bounds, and trade-offs among competing objectives. We study three mechanisms for supplying this knowledge to an LLM warm-start generator:

\begin{enumerate}
    \item \emph{Staged model reasoning and self-verification:} Prompt decomposition breaks complex reasoning into sequential subtasks \citep{khot2022decomposed}, while Self-Refine uses model-generated feedback to revise initial outputs \citep{madaan2023selfrefine}. AMP applies both ideas to warm starts (Section~\ref{sec:amp}).
    \item \emph{External documentation and statistical scouting:} Retrieval-augmented generation (RAG) grounds model outputs in external text corpora \citep{lewis2020retrieval}. HKMA adds retrieved documentation, or a summary of TPE scouting, to the prompt (Section~\ref{sec:hkma}).
    % \item \emph{Empirical feature ranking:} High-dimensional configuration spaces often contain many low-impact parameters \citep{xu2015hey}. DAPR shows the LLM the most important features first, then adds the rest step by step (Section~\ref{sec:dapr}).
    \item \emph{Human expert review:} Although LLMs can match human agreement on certain software annotation tasks \citep{ahmed2025llmsreplacemanualannotation}, unverified model beliefs can still encode flawed domain assumptions. H-DKP asks human experts to check the LLM's domain rules before optimization starts (Section~\ref{sec:hdkp}).
\end{enumerate}

\subsection{Tabular benchmarks}
We evaluate all methods using the MOOT repository \citep{mootrepo}, which curates tabular multi-objective optimization datasets from empirical software engineering studies, including measured system configurations such as those analyzed in PromiseTune \citep{chen2026promisetune} and software process simulations \citep{port2008pom3,menzies2005xomo}. Tabular benchmarks map pre-recorded configurations to measured objective values, enabling reproducible comparisons across many trials without re-executing costly system workloads. Because an LLM can propose arbitrary continuous or categorical feature combinations, each generated candidate is mapped to the nearest observed row in the benchmark table. Studies on surrogate and tabular benchmarks, such as YAHPO Gym \citep{pmlr-v188-pfisterer22a} and NAS benchmarks \citep{zela2022surrogatenasbenchmarksgoing}, note that discrete table resolution can influence evaluation when proposals fall far from recorded rows. We account for this representation in our threats to validity and inspect proposal-to-row distances when constructing follow-up queries in our human expert study.

\subsection{Humans in the optimization loop}\label{sec:hitl}
Interactive search-based software engineering asks humans to guide or judge the search \citep{ramirez2019interaction}. A long-standing problem with this approach is user fatigue: people tire quickly, so an interactive optimizer can ask them for only a small number of judgments \citep{takagi2001interactive}. Knowledge acquisition has the same problem. Experts are slow to label examples \citep{valerdi2010heuristics,lustosa2024learning}, and in the 1980s, knowledge engineers expected only 5--10 rules per expert per day \citep{menzies1992expert}. Hence, human effort must be spent where it matters most. In H-DKP, experts do not label configurations; instead, they review a short list of rules proposed by the LLM across a few brief email exchanges (Sections~\ref{sec:human-study} and~\ref{sec:rq3}).

\section{Methods}\label{sec:methods}
Different optimization scenarios provide access to different sources of information.
Hence our three knowledge-guided approaches elicit and enforce this knowledge through differing mechanisms: AMP relies on staged model reasoning and self-verification; HKMA relies on preliminary statistical scouting and external documentation; and H-DKP incorporates asynchronous human expert review.

\subsection{Common warm-start generation and sequential search}
We say that $\mathcal{D}=\{(x_i,y_i)\}_{i=1}^{N}$ denotes a benchmark dataset where each configuration $x_i$ comprises $p$ input features and each outcome $y_i$ comprises $m$ objective values with defined optimization directions (minimization or maximization).
All single-session LLM warm-start methods (BS-LLM, AMP, HKMA, and H-DKP) share a common initialization and downstream optimization interface. Each trial begins by sampling $B_{\mathrm{initial}}=4$ observed configurations from the benchmark dataset $\mathcal{D}$ and partitioning them by multi-objective quality into the best-performing observations (\emph{best}) and the remaining observations (\emph{rest}). Supplying both subsets gives the language model and the downstream sequential optimizer explicit examples of promising and unpromising parameter regions.

Given the task metadata $M$ (feature names, types, value ranges, and objective directions), the labeled best/rest examples $S$, and any method-specific domain knowledge, the LLM is prompted to generate $B_{\mathrm{generated}}=4$ feature-only candidate configurations: two expected to perform well and two expected to perform poorly. Requesting contrastive proposals encourages the model to span both high-performing and failure regions so the sequential surrogate can establish an informative decision boundary early in the search. Generated candidates are validated against the dataset schema for complete fields and valid data types, mapped to their nearest observed rows in $\mathcal{D}$ in normalized feature space, and evaluated by retrieving their objective values.

Following warm-start evaluation, sequential model-based optimization (SMO) iteratively selects and labels the remaining configurations up to the total budget $B_{\mathrm{total}}=24$. At the start of SMO and after each subsequent evaluation, all observed configurations (the initial seeds plus the evaluated warm-start rows) are re-partitioned into \emph{best} and \emph{rest} subsets using their actual objective values rather than the LLM's intended better/poorer polarity. Using the same lightweight active-learning core as EZR \citep{menzies2026can}, SMO fits a two-class Naive Bayes surrogate with Gaussian likelihoods over the observed \emph{best} and \emph{rest} partitions and selects the next configuration via its exploit acquisition policy. For BS-LLM, AMP, HKMA-rag, and H-DKP, SMO executes $B_{\mathrm{search}}=16$ sequential steps ($4 + 4 + 16 = 24$). For HKMA-scout and HKMA-combined, which allocate $B_{\mathrm{scout}}=10$ evaluations to preliminary statistical scouting prior to LLM generation, SMO executes $B_{\mathrm{search}}=6$ sequential steps ($4 + 10 + 4 + 6 = 24$).

\subsection{Adaptive Multi-Stage Prompting (AMP)}\label{sec:amp}
Direct few-shot prompting asks the LLM to infer parameter interactions and output numerical configurations in a single generation pass. In complex spaces, models frequently hallucinate parameter settings that violate basic domain constraints. Adaptive Multi-Stage Prompting (AMP) decomposes warm-start generation into sequential reasoning stages \citep{khot2022decomposed,madaan2023selfrefine} to test whether explicit constraint proposals and per-candidate verification improve configuration quality without requiring external data sources.

\begin{widetable}[t]
\caption{AMP ablation conditions and nominal LLM calls required for a four-candidate warm start. Validation in AMP-4 is executed once per generated candidate.}\label{tab:amp}
\centering\small
\begin{tabular}{lcccr}
Condition & Analysis & Constraints & Validation & Nominal calls\\
\midrule
AMP-2 & Yes & No & No & 2\\
AMP-3 & Yes & Yes & No & 3\\
AMP-4 & Yes & Yes & Yes & 7
\end{tabular}
\end{widetable}

Table~\ref{tab:amp} and Algorithm~\ref{alg:amp} define the three stage ablations \citep{registeredreport2026}. In the first stage, the model analyzes the task metadata $M$ and initial examples $S$ to summarize likely feature-objective trade-offs. AMP-2 passes this analysis directly to the candidate generation prompt (2 nominal LLM calls). AMP-3 introduces an intermediate stage that prompts the model to formulate explicit feasibility constraints and parameter rules $C$ from its analysis before generating the four candidates (3 nominal LLM calls). However, an LLM may articulate valid constraints in natural language yet still violate them when generating multi-dimensional vectors. AMP-4 addresses this gap by adding a dedicated validation stage: each of the four generated candidates is checked individually against the discovered constraints $C$, and any candidate flagged as violating a constraint is revised before schema verification (7 nominal calls). All three variants operate within the same 24-label evaluation budget.

\begin{algorithm}[tbp]
\caption{AMP generation and stage ablations}\label{alg:amp}
\small
\begin{algorithmic}[1]
\REQUIRE Metadata $M$, initial best/rest examples $S$, mode $m\in\{2,3,4\}$
\ENSURE Four feature-only warm-start candidates $E$
\STATE $A\gets\mathrm{LLM}(M,S,\text{analyze features and trade-offs})$
\STATE $C\gets\varnothing$
\IF{$m\geq3$}
    \STATE $C\gets\mathrm{LLM}(A,\text{discover constraints})$
\ENDIF
\STATE $E\gets\mathrm{LLM}(M,S,A,C,\text{generate two better and two poorer candidates})$
\IF{$m=4$}
    \FOR{each candidate $e\in E$}
        \STATE $v\gets\mathrm{LLM}(C,e,\text{validate and propose a revision if needed})$
        \IF{$v$ marks $e$ invalid and supplies revised features}
            \STATE Replace $e$ with the proposed revision
        \ENDIF
    \ENDFOR
\ENDIF
\STATE Apply common schema checks; reject malformed final sets
\RETURN $E$
\end{algorithmic}
\end{algorithm}
\FloatBarrier

\subsection{Hybrid Knowledge-Model Approach (HKMA)}\label{sec:hkma}
Rather than relying solely on internal model reasoning, the Hybrid Knowledge-Model Approach (HKMA) supplies external context derived from preliminary statistical scouting, technical documentation, or their combination (Algorithm~\ref{alg:hkma}) \citep{registeredreport2026}.

We evaluate three variants under the fixed 24-label budget:
\begin{itemize}
    \item \emph{HKMA-scout:} A Tree-structured Parzen Estimator (TPE) \citep{NIPS2011_86e8f7ab} selects $B_{\mathrm{scout}}=10$ distinct observed configurations from $\mathcal{D}$. The LLM is prompted to summarize empirical parameter patterns $P$ that distinguish the higher-performing and lower-performing scouted rows, and this summary is injected into the warm-start generation prompt alongside updated best/rest examples. Because the 10 scouting evaluations are part of the 24-label budget, HKMA-scout trades fewer downstream SMO iterations ($B_{\mathrm{search}}=6$) for a richer empirical summary prior to generation.
    \item \emph{HKMA-rag:} Using task and feature descriptions as a query $q$, a retrieval-augmented generation pipeline \citep{lewis2020retrieval} extracts relevant passages $R$ from a curated index $I$ of domain documentation and system manuals. The retrieved passages are supplied directly to the warm-start prompt without consuming any scouting labels, preserving $B_{\mathrm{search}}=16$ SMO steps.
    \item \emph{HKMA-combined:} TPE scouting first acquires $B_{\mathrm{scout}}=10$ observations and generates the empirical summary $P$. That summary is combined with the task metadata to form an informed retrieval query $q$, and both $P$ and the retrieved passages $R$ are provided to the generation prompt, followed by $B_{\mathrm{search}}=6$ SMO steps. Comparing HKMA-combined against HKMA-scout and HKMA-rag tests whether combining empirical scouting with documentation yields additive gains.
\end{itemize}

\begin{algorithm}[tbp]
\caption{HKMA generation under alternative knowledge conditions}\label{alg:hkma}
\small
\begin{algorithmic}[1]
\REQUIRE Metadata $M$, initial examples $S$, document index $I$, condition $m$
\ENSURE Four feature-only candidates $E$ and recorded scouting evaluations
\STATE $P\gets\varnothing$; $R\gets\varnothing$
\IF{$m\in\{\text{scout},\text{combined}\}$}
    \STATE $S_{\mathrm{scout}}\gets$ ten distinct observed rows selected by TPE scouting
    \STATE Evaluate these rows and charge ten evaluations against the total labeling budget
    \STATE $P\gets\mathrm{LLM}(S_{\mathrm{scout}},\text{summarize observed best/rest differences})$
    \STATE $S\gets$ selected best/rest examples from $S\cup S_{\mathrm{scout}}$
\ENDIF
\IF{$m\in\{\text{rag},\text{combined}\}$}
    \STATE $q\gets$ feature/objective metadata, plus $P$ when available
    \STATE $R\gets\mathrm{Retrieve}(I,q)$
\ENDIF
\STATE $E\gets\mathrm{LLM}(M,S,P,R,\text{generate two better and two poorer candidates})$
\STATE Apply common final-set schema checks
\RETURN $E$ and recorded scouting evaluations
\end{algorithmic}
\end{algorithm}
\FloatBarrier

\subsection{Human-in-the-Loop Domain Knowledge Prompting (H-DKP)}\label{sec:hdkp}
Automated prompting and document retrieval can still introduce spurious correlations or misinterpret domain-specific parameters. Human-in-the-Loop Domain Knowledge Prompting (H-DKP) incorporates asynchronous expert review to filter, qualify, and refine LLM-generated domain claims before freezing the resulting knowledge for optimization (Algorithm~\ref{alg:hdkp}) \citep{registeredreport2026}. Participant demographics, ethics approval, and measured effort across the nine completed expert tracks are detailed in Section~\ref{sec:human-study}.

As shown in Algorithm~\ref{alg:hdkp}, H-DKP separates offline knowledge acquisition from online optimization. First, using only task metadata $M$ and source documentation $C$, the LLM is prompted at temperature 0 to propose at most ten domain claims $K_0$, each stating a hypothesized feature-objective relationship, rationale, and confidence score. After automated screening to verify that all referenced feature names exist in $M$, a human specialist classifies each claim as \textsc{valid}, \textsc{modify}, \textsc{invalid}, or \textsc{unsure}; only claims marked \textsc{valid} or supplied with explicit replacement conditions (\textsc{modify}) enter the active rule set $K$. Across subsequent asynchronous contacts ($T \le 10$), the LLM generates candidate configurations from $K$ and presents one diagnostic question per exchange (either a failure-diagnosis query when an expected-better candidate underperforms the initial seeds, or a profile-plausibility query when the nearest benchmark row is too distant), updating $K$ only with explicit expert rules. Once frozen, $K$ conditions 20 independent warm-start trials under the standard 24-label budget.

\begin{algorithm}[tbp]
\caption{H-DKP with asynchronous expert review}\label{alg:hdkp}
\small
\begin{algorithmic}[1]
\REQUIRE Source context $C$, metadata $M$, expert $H$, stopping rule $T\leq10$
\ENSURE Frozen knowledge $K$ and 20 independent four-candidate sets
\STATE Propose initial claims $K_0\gets\mathrm{LLM}(C,M)$ without observed rows
\STATE Screen claim schema and send eligible claims to $H$
\STATE $K\gets$ explicitly accepted or revised claims from the reply
\FOR{each subsequent expert exchange until the stopping rule is met}
    \STATE $E\gets\mathrm{LLM}(M,K)$
    \STATE Validate proposals and evaluate mapped observed rows
    \IF{an interpretable failure candidate is available}
        \STATE Prepare a failure-diagnosis question using current knowledge $K$
    \ELSE
        \STATE Prepare a profile-plausibility question for an unqueried candidate
    \ENDIF
    \STATE Send question to $H$ and await reply
    \STATE Update $K$ only with explicit accepted wording and conditions
\ENDFOR
\STATE Freeze $K$ separately for this expert track
\FOR{$r=1$ to $20$}
    \STATE $E_r\gets\mathrm{LLM}(M,K,\text{trial examples}_r)$
    \STATE Validate and score the four proposals, then run $B_{\mathrm{search}}=16$ SMO steps
\ENDFOR
\RETURN $K,\{E_r\}_{r=1}^{20}$
\end{algorithmic}
\end{algorithm}
\FloatBarrier

\subsection{Comparison baselines}\label{sec:baselines}
We compare the knowledge-guided methods against three model-free baselines and two unguided LLM baselines:
\begin{itemize}
    \item \emph{Random search:} Uniformly samples $B_{\mathrm{search}}=24$ distinct observed configurations from $\mathcal{D}$ without replacement and returns the best configuration observed \citep{bergstra2012random}.
    \item \emph{EZR:} A lightweight active-learning optimizer \citep{menzies2026can} that initializes with $B_{\mathrm{initial}}=4$ randomly selected configurations and iteratively selects $B_{\mathrm{search}}=20$ additional configurations using its exploit policy, yielding 24 distinct configuration labels per trial.
    \item \emph{Recursive Random Projection (RRP):} A sampling-based baseline that recursively partitions the configuration space using random projections, evaluating representative boundary instances at each split \citep{chen2018sway}. Its evaluation count depends on dataset size and recursion depth.
    \item \emph{Direct few-shot prompting (BS-LLM):} Following \citet{senthilkumar2024largelanguagemodelsimprove}, BS-LLM prompts the language model with task metadata $M$ and $B_{\mathrm{initial}}=4$ initial best/rest examples to propose $B_{\mathrm{generated}}=4$ warm-start candidates in a single call, followed by $B_{\mathrm{search}}=16$ SMO evaluations ($B_{\mathrm{total}}=24$).
    % \item \emph{Full SynthCore:} We run the Diversity ensemble routine from the SynthCore reference implementation \citep{senthilkumar2026synthcore}. It selects $B_{\mathrm{initial}}=16$ initial representatives via recursive partitioning and executes 15 independent few-shot generation sessions, each proposing 4 candidates mapped to observed rows in $\mathcal{D}$, before selecting the top configuration across all sessions. This procedure issues $16 + 15 \times 4 = 76$ nominal evaluation requests per trial. Because multiple sessions can map to the same nearest-neighbor row in $\mathcal{D}$ (or return fewer than four valid candidates), full SynthCore evaluates an average of 55 distinct configuration labels per trial.
    \item \emph{Three-session budget-matched SynthCore (SynthCore-28):} We run the Diversity ensemble routine from SynthCore \citep{senthilkumar2026synthcore}. It selects $B_{\mathrm{initial}}=16$ initial representatives via recursive partitioning and executes 15 independent few-shot generation sessions, each proposing 4 candidates mapped to observed rows in $\mathcal{D}$, before selecting the top configuration across all sessions. This procedure issues $16 + 15 \times 4 = 76$ nominal evaluation requests per trial. Because full SynthCore consumes more than double our 24-label budget, we evaluate a budget-matched restriction, SynthCore-28, constructed from the 16 initial representatives and three generation sessions of each trial ($16 + 3 \times 4 = 28$ nominal evaluation requests, averaging 26.31 distinct configuration labels). Comparing SynthCore-28 against BS-LLM and AMP-4 isolates whether multi-session ensembling remains competitive when restricted to roughly two dozen configuration labels.
\end{itemize}

\section{Experimental setup}\label{sec:design}

Our four research questions (Section~\ref{sec:introduction}) use different parts of this study. RQ1 and RQ2 evaluate the ten automated 24-label methods across 49 MOOT datasets. RQ3 evaluates H-DKP along with ten baselines across eight datasets with nine independent expert tracks. RQ4 examines labeling, computational, and human effort costs across all methods, along with a budget-matched three-session SynthCore baseline (SynthCore-28) to test whether multi-session ensembling remains competitive under small labeling budgets. Across all research questions, we evaluate the final configuration returned by each method using Chebyshev distance to the ideal point.

\subsection{Datasets and dimensionality}
To assess how methods scale with parameter space complexity (RQ2), we partition the 49 MOOT datasets by input feature count $p$ into three tiers: low dimensionality ($p < 6$, 15 datasets), medium dimensionality ($6 \le p \le 11$, 17 datasets), and high dimensionality ($p > 11$, 17 datasets). The full 49-dataset corpus spans measured software configurations (such as database, compiler, and learner tuning), software process simulations, and general optimization tasks that broaden the range of feature interactions and objective trade-offs.

For more details on this data, see the MOOT documentation~\citep{mootrepo}.

\subsection{The human-in-the-loop study}\label{sec:human-study}
Evaluating H-DKP (RQ3) requires human specialists to audit technical parameter relationships. Our protocol was approved by the North Carolina State University Institutional Review Board (protocol \#28893). We recruited ten participants via email based on relevant software engineering, systems, or quantitative domain experience. Each candidate received an informed consent document by email and confirmed participation by email reply prior to the first round of questions. Participation was voluntary and uncompensated. Nine participants completed the protocol and one withdrew prior to completion (our records do not log the withdrawn participant's assigned dataset), yielding nine completed expert tracks across eight datasets: two low-dimensional, four medium-dimensional, and two high-dimensional tasks (Table~\ref{tab:expert-cohort}).

\begin{widetable}[t]
\caption{Participants and datasets in the H-DKP study. Each track represents one specialist's asynchronous rule review followed by 20 optimization repeats using the same rules. Columns $p$ and $m$ give input features and objectives; Tier denotes dimensionality (Low $p<6$, Med $6\le p\le 11$, High $p>11$); Contacts gives completed exchanges; Prep.\ gives benchmark accesses during offline follow-up question generation (Gen.\ exch.).}\label{tab:expert-cohort}
\centering\scriptsize
\setlength{\tabcolsep}{3pt}
\begin{adjustbox}{width=\textwidth}
\begin{tabular}{llrrllrrr}
ID & Dataset & $p$ & $m$ & Tier & Role / expertise basis & Yrs & Contacts & Prep.\ (Gen.\ exch.)\\
\midrule
E1 & Auto93 & 4 & 3 & Low & Mechanical engineer & 4 & 9 & 42 (7)\\
E2 & Project health & 5 & 3 & Low & Industrial repository maintainer & 2 & 6 & 12 (2)\\
E3 & A2C CartPole & 7 & 4 & Med & Software engineer & 4 & 8 & 36 (6)\\
E4 & PostgreSQL & 9 & 1 & Med & Software engineer & 4 & 8 & 18 (3)\\
E5 & POM3a (Track 1) & 9 & 3 & Med & Published articles in area & 4 & 10 & 12 (2)\\
E6 & POM3a (Track 2) & 9 & 3 & Med & Published articles in area & 30 & 10 & 18 (3)\\
E7 & LLVM & 10 & 1 & Med & Researcher in relevant area & 2 & 10 & 24 (4)\\
E8 & Life expectancy & 20 & 2 & High & Biostatistician in public health & 4 & 10 & 48 (8)\\
E9 & XOMO Flight & 23 & 4 & High & Researcher in relevant area & 2 & 10 & 6 (1)\\
\midrule
\multicolumn{2}{l}{Total (8 datasets, 9 tracks)} & & & & & & 81 & 216 (36)
\end{tabular}
\end{adjustbox}
\end{widetable}

Table~\ref{tab:expert-cohort} summarizes the nine specialists (\texttt{E1}--\texttt{E9}), whose domain tenure ranged from 2 to 30 years. Within this cohort, the POM3 family models requirements prioritization under shifting stakeholder constraints \citep{port2008pom3}, and XOMO Flight simulates autonomous software development risk and effort \citep{menzies2005xomo}. Two independent specialists (\texttt{E5} and \texttt{E6}) evaluated POM3a starting from the same initial LLM claims to test whether different experts produce consistent optimization outcomes on the same task.

As formalized in Section~\ref{sec:hdkp} (Algorithm~\ref{alg:hdkp}), each track proceeded in two phases. In Phase~A (offline knowledge acquisition), the expert reviewed up to ten LLM claims (\textsc{valid}, \textsc{modify}, \textsc{invalid}, or \textsc{unsure}) and then answered one diagnostic question per follow-up email (either a failure diagnosis or a plausibility check) before freezing the knowledge base $K$ after 6--10 exchanges. Across the nine tracks, 36 follow-up exchanges involved generating and matching intermediate configurations against the benchmark table; under the nominal protocol of four seed rows and two candidate matches per generation exchange, offline preparation incurred an estimated 6--48 nominal table accesses per track (216 across all nine tracks; Table~\ref{tab:expert-cohort}). Because seed rows and nearest-neighbor matches can repeat across exchanges, distinct offline labels are bounded above by these nominal counts. In Phase~B (online optimization), the frozen rule set $K$ guided 20 independent trials, each strictly capped at the 24-label online budget ($B_{\mathrm{initial}}=4, B_{\mathrm{generated}}=4, B_{\mathrm{search}}=16$). Crucially, these offline accesses do not pass observed configurations or objective scores into Phase B: all intermediate configurations evaluated during Phase A are discarded after question generation, and only qualitative, natural-language directional rules explicitly confirmed by the specialist enter the frozen knowledge base $K$.

Across all nine tracks, Phase~A totaled 81 elicitation exchanges (6--10 per track, plus one recruitment and one thank-you email), completed over 5--17 calendar days per specialist. Because the first contact presented up to ten initial claims at once, initial calibration replies were longest (at most 19 lines), whereas subsequent single-question exchanges required only 1--3 lines per reply. Box~1 illustrates two claim-review exchanges.

\begin{tcolorbox}[enhanced, colback=gray!4, colframe=gray!60, boxrule=0.6pt, sharp corners, left=6pt, right=6pt, top=4pt, bottom=4pt, title={\footnotesize\bfseries Box 1: Sample expert claim-review exchanges from H-DKP (Phase A)}, coltitle=black, colbacktitle=gray!20]
\footnotesize
\textbf{Example A -- PostgreSQL (Rule R08, rejected):}
\begin{itemize}
\item \emph{Initial LLM claim:} ``If \texttt{fsync} is set to off, \texttt{fullPageWrites} should be kept on to mitigate increased risk of page corruption.''
\item \emph{Expert \texttt{E4} verdict (\textsc{invalid}):} ``Doesn't make sense to keep FullPageWrites on if Fsync is already off if mitigation is the goal.''
\item \emph{Resulting rule treatment:} R08 was removed from the frozen knowledge base (without inferring an unverified opposite rule).
\end{itemize}
\smallskip
\textbf{Example B -- POM3a (Rule R07, qualified vs.\ retained):}
\begin{itemize}
\item \emph{Initial LLM claim:} ``Higher \texttt{Culture} values (more requirements change) are likely to increase both simulated project cost (\texttt{Cost-}) and team idleness (\texttt{Idle-}).''
\item \emph{Expert \texttt{E5} verdict (\textsc{modify}):} ``Really depends on the change. If by change you mean tiny adoptable changes, the statement does not apply.''
\item \emph{Resulting frozen rule (Track 1):} ``If more requirements change, project cost and team idle time are likely to increase, subject to the scope of the change and existing foundations.'' (By contrast, Expert \texttt{E6} in Track 2 classified R07 as \textsc{valid} and retained the broader rule.)
\end{itemize}
\end{tcolorbox}

\subsection{Budgets and model configuration}\label{sec:analysis}
Figure~\ref{fig:warm-start-overview} shows the study workflow. In the single-session LLM methods (BS-LLM, AMP, HKMA, and H-DKP), each online trial begins with $B_{\mathrm{initial}}=4$ labeled examples partitioned into best and rest sets. The knowledge-construction procedure guides the LLM to propose $B_{\mathrm{generated}}=4$ candidate configurations (two expected to perform well and two expected to perform poorly), which are mapped to their nearest observed rows in $\mathcal{D}$ and labeled. A sequential model-based optimizer (SMO) then selects and labels the remaining configurations up to a fixed total budget of $B_{\mathrm{total}}=24$ evaluations ($B_{\mathrm{search}}=16$ for BS-LLM, AMP, HKMA-rag, and H-DKP; $B_{\mathrm{scout}}=10$ and $B_{\mathrm{search}}=6$ for HKMA-scout and HKMA-combined). For the budget-matched ensemble sensitivity baseline, SynthCore-28 is configured as described in Section~\ref{sec:baselines}. We run 20 independent trials per method and dataset across all experiments.

All LLM-guided methods use GPT-OSS-120B \citep{agarwal2025gpt}\footnote{We select GPT-OSS-120B for three practical and methodological reasons: (1)~as an open-weights model, its fixed parameter snapshot supports long-term scientific reproducibility without the risk of silent model updates or endpoint deprecations common to proprietary frontier APIs \citep{sallou2024breaking}; (2)~it is widely accessible through standard inference providers such as OpenRouter without requiring dedicated local GPU clusters; and (3)~its low per-token pricing makes large-scale empirical benchmarking across 49 datasets (with 20 repetitions) and multiple multi-stage prompting ablations economically feasible.} with a decoding temperature of 0.7, top-$p$ of 1.0, and a maximum of 4,000 output tokens per call during candidate generation. Initial domain claims in H-DKP are elicited at temperature 0 to ensure deterministic baseline beliefs for expert review (Section~\ref{sec:human-study}). For every completed trial, we log the number of LLM API calls, input and output tokens, monetary API charges, and elapsed wall-clock time.

To ensure a fair comparison, and following principles of experimental algorithmics \citep{mcgeoch2012guide}, we account for all online configuration evaluations consumed across every stage of an optimization run:
\begin{equation}
B_{\mathrm{total}}=B_{\mathrm{initial}}+B_{\mathrm{scout}}+B_{\mathrm{generated}}+B_{\mathrm{search}}.
\label{eq:budget}
\end{equation}
Here, $B_{\mathrm{initial}}$ denotes initial seed observations, $B_{\mathrm{scout}}$ denotes preliminary TPE scouting evaluations, $B_{\mathrm{generated}}$ denotes warm-start proposals mapped to benchmark rows, and $B_{\mathrm{search}}$ denotes subsequent sequential search steps. Because mapping multiple generated candidates to their nearest neighbors in a finite table can select the same row more than once, we also track the number of distinct observed configurations evaluated per trial. Table~\ref{tab:budget-breakdown} summarizes the online evaluation budgets across all methods; for H-DKP, one-time offline preparation accesses (6--48 nominal accesses per track; Table~\ref{tab:expert-cohort}) are incurred once prior to freezing the knowledge base and are separate from the 24-label per-trial budget.

\begin{widetable}[t]
\caption{Configuration evaluation budgets across methods, decomposed by Equation~(\ref{eq:budget}). Nominal $B_{\mathrm{total}}$ gives total evaluation requests per trial; Distinct gives the mean number of unique benchmark configurations labeled.}\label{tab:budget-breakdown}
\centering\scriptsize
\setlength{\tabcolsep}{4.5pt}
\begin{adjustbox}{width=\textwidth}
\begin{threeparttable}
\begin{tabular}{lrrrrrr}
Method & $B_{\mathrm{initial}}$ & $B_{\mathrm{scout}}$ & $B_{\mathrm{gen.}}$ & $B_{\mathrm{search}}$ & Nominal $B_{\mathrm{total}}$ & Distinct\\
\midrule
Random & 0 & 0 & 0 & 24 & 24 & 24\\
EZR & 4 & 0 & 0 & 20 & 24 & 24\\
RRP & \multicolumn{4}{c}{Recursive partitioning by dataset size} & Dataset-dep. & Dataset-dep.\\
BS-LLM & 4 & 0 & 4 & 16 & 24 & $\le 24$\\
AMP-2 / 3 / 4 & 4 & 0 & 4 & 16 & 24 & $\le 24$\\
HKMA-rag & 4 & 0 & 4 & 16 & 24 & $\le 24$\\
HKMA-scout & 4 & 10 & 4 & 6 & 24 & $\le 24$\\
HKMA-combined & 4 & 10 & 4 & 6 & 24 & $\le 24$\\
H-DKP\tnote{a} & 4 & 0 & 4 & 16 & 24 & $\le 24$\\
\midrule
SynthCore-28 & 16 & 0 & 12 & 0 & 28 & 26.31
\end{tabular}
\end{threeparttable}
\end{adjustbox}
\footnotetext[a]{Excludes one-time offline rule preparation (6--10 expert contacts and 6--48 nominal benchmark accesses per track; Section~\ref{sec:human-study}).}
\end{widetable}

\subsection{Final configuration quality}\label{sec:performance}
We evaluate each method by the multi-objective quality of the final configuration $x$ it returns. For each objective $j \in \{1,\dots,m\}$, let $a_j$ and $b_j$ denote the minimum and maximum observed values across the reference dataset $\mathcal{D}$. Each nonconstant objective is min-max normalized to the unit interval:
\begin{equation}
u_j(x)=\frac{y_j(x)-a_j}{b_j-a_j}.
\end{equation}
Let $z_j$ denote the ideal normalized value for objective $j$, where $z_j=0$ for minimization objectives and $z_j=1$ for maximization objectives. We measure the quality of configuration $x$ using its Chebyshev ($L_\infty$) distance to the ideal point $z$:
\begin{equation} 
D(x)=\max_{1\leq j\leq m}|u_j(x)-z_j|.
\label{eq:chebyshev}
\end{equation}
Lower values of $D(x)$ indicate better performance. Unlike averaging metrics that allow poor performance on one objective to be masked by gains on another, Chebyshev distance penalizes the worst normalized shortfall across all $m$ objectives \citep{Li22}. Because the ideal point $z$ combines the best observed value of each individual objective in $\mathcal{D}$, it typically lies outside the feasible Pareto front, meaning $D(x) > 0$ even for Pareto-optimal configurations.

During optimization, SMO, EZR, and SynthCore-28 rank candidate configurations internally using their native acquisition heuristic, Euclidean distance to heaven, $\sqrt{m^{-1}\sum_{j=1}^{m}|u_j(x)-z_j|^2}$. We preserve each optimizer's internal $L_2$ ranking unmodified so that baselines operate as designed in prior work \citep{menzies2026can,senthilkumar2026synthcore}, while scoring the final configuration returned by every method with Chebyshev distance $D(x)$ in Equation~(\ref{eq:chebyshev}) as an independent, external audit.

\subsection{Statistical comparisons and grouping}\label{sec:combined-analysis}
Because optimization quality depends on how many configurations an algorithm is permitted to evaluate, comparing a 24-label method directly against a multi-session ensemble that consumes over four dozen labels confounds knowledge quality with sample size. We therefore structure our statistical comparisons around three scopes restricted to roughly two dozen evaluations:

\begin{enumerate}
    \item \emph{Primary benchmark comparison (RQ1 and RQ2):} Compares the ten automated 24-label methods (Random, EZR, RRP, BS-LLM, AMP-2, AMP-3, AMP-4, HKMA-scout, HKMA-rag, and HKMA-combined) across all 49 datasets.
    \item \emph{Expert-study comparison (RQ3):} Compares H-DKP against the same ten automated methods across the eight expert-study datasets, analyzing each of the two POM3a expert tracks independently.
    \item \emph{Budget-matched ensemble sensitivity comparison (RQ4):} Evaluates the three-session SynthCore-28 baseline (28 nominal requests; 26.31 mean distinct labels) against the ten 24-label methods across all 49 datasets, alongside an analysis of API token costs, runtime, and human preparation effort.
\end{enumerate}

Within each dataset and comparison scope, we rank methods using the Scott--Knott clustering procedure \citep{scott1974,mittas2012ranking}. Methods are sorted by their mean Chebyshev distance across the 20 trials and recursively partitioned into disjoint groups to maximize the between-group sum of squares. A candidate split is accepted only if the two resulting groups exhibit both a non-negligible effect size via Cliff's delta ($|\delta| \ge 0.147$) \citep{Cliff1993DominanceSO,romano2006appropriate} and statistical significance ($p < 0.05$) under a studentized bootstrap test \citep{tibshirani1993introduction} with 512 resamples under a common-mean null and an add-one pseudocount correction \citep{phipson2016permutation}.\footnote{All bootstrap tests use independent pseudorandom streams derived from base seed 20261016.}

Methods assigned to Group 1 (the lowest-scoring cluster) on a given dataset are statistically indistinguishable winners for that task. Because Scott--Knott splits depend on the set of candidate distributions being partitioned, group memberships are computed separately for each comparison scope. Across datasets, we report both the number of tasks on which each method enters Group 1 (best-group membership) and the equal-weighted mean Chebyshev distance across the benchmark corpus.

\subsection{Ablation and expert-rule analysis}
To understand why specific knowledge strategies succeed or fail, our experimental design includes controlled ablations alongside the human expert audit. Comparing BS-LLM, AMP-2, AMP-3, and AMP-4 under an identical 24-label budget isolates the incremental effect of task analysis, constraint discovery, and per-candidate constraint validation. Similarly, comparing HKMA-scout, HKMA-rag, and HKMA-combined isolates the contribution of empirical scouting summaries versus retrieved documentation. For H-DKP, we record the expert classification (\textsc{valid}, \textsc{modify}, \textsc{invalid}, or \textsc{unsure}) for every initial LLM claim, as well as the rules added or qualified during follow-up exchanges, allowing us to relate rule curation patterns to downstream optimization medians.

\section{Results}\label{sec:results}
Our experiments follow one design principle:   human time is
the scarce resource, so we use it last and only where it matters
most. Hence, our experiments run in two stages:
\begin{itemize}
\item \textbf{Make elaborate use of cheap CPU time:} We run 10 optimizers
  on 49 datasets and keep the best one. As seen below,
  the automatic analysis of RQ1 and RQ2 selected AMP-4.
\item \textbf{Take more care with expensive human time:} RQ3 compares our human-in-the-loop method (H-DKP) against the ten automated 24-label methods across eight datasets (nine expert tracks), and RQ4 evaluates the label, compute, and human effort costs across methods.
\end{itemize}

\subsection{RQ1: Automated warm starts with 24 labels}\label{sec:rq1}
\noindent{\bf RQ1:} {\bf With only 24 labels, can domain knowledge help an LLM propose better initial configurations?}

If RQ1 is true, then we would expect all our optimizers to perform better than some simpler baseline (i.e. BS-LLM's single-shot prompting that uses no background domain knowledge).

As shown in Table~\ref{tab:primary-low-label}, this expectation was not met:
\begin{itemize}
\item
BS-LLM was as good as, or better than, nearly everything else.
\item
The sole exception was AMP-4.
Among all compared approaches, AMP-4 achieves both the broadest statistical win rate, entering the best Scott--Knott group (Group 1) on 44 of 49 datasets (89.8\%), and the lowest equal-weighted mean Chebyshev distance ($0.1862$).
\end{itemize}

% \begin{table}[t]

% \caption{Final Chebyshev distance ($D$) and Scott--Knott Group 1 (best-group) membership counts across the 49 datasets for the ten scarce-label methods. Lower $D$ values are better. Dimensionality tiers contain 15 low ($p<6$), 17 medium ($6\le p\le 11$), and 17 high ($p>11$) datasets.}\label{tab:primary-low-label}
% \scriptsize

% \setlength{\tabcolsep}{4pt}
% \begin{tabular}{lrrrrr}
%        &               &\multicolumn{3}{c}{\#configuration options}&\\ \cmidrule(lr){3-5}
% Method & Best /49 (\%) & Low /15 & Medium /17 & High /17 & Mean $D$\\
% \midrule
% AMP-4 & 44 (89.8\%) & 13 & 15 & 16 & 0.1862\\
% BS-LLM & 36 (73.5\%) & 11 & 12 & 13 & 0.1959\\
% AMP-2 & 36 (73.5\%) & 11 & 12 & 13 & 0.1965\\
% AMP-3 & 36 (73.5\%) & 10 & 14 & 12 & 0.1987\\
% EZR & 34 (69.4\%) & 10 & 13 & 11 & 0.1953\\
% HKMA-scout & 33 (67.3\%) & 11 & 10 & 12 & 0.1956\\
% HKMA-combined & 33 (67.3\%) & 11 & 11 & 11 & 0.1963\\
% HKMA-rag & 33 (67.3\%) & 12 & 12 & 9 & 0.1971\\
% Random & 23 (46.9\%) & 8 & 9 & 6 & 0.2226\\
% RRP & 13 (26.5\%) & 3 & 4 & 6 & 0.2675
% \end{tabular}
% \end{table}

\begin{widetable}[t]
\caption{Final Chebyshev distance ($D$) and Scott--Knott Group 1 (best-group) membership counts across the 49 datasets for the ten automated 24-label methods. Lower $D$ values are better. Dimensionality tiers contain 15 low ($p<6$), 17 medium ($6\le p\le 11$), and 17 high ($p>11$) datasets.}
\label{tab:primary-low-label}

\scriptsize
\setlength{\tabcolsep}{4pt}

~\\

\begin{tabular}{lrrrr|ll}
       & & \multicolumn{3}{c}{\#configuration options} & Mean $D$ \\\cmidrule(lr){3-5}
Method & Best /49 (\%) & Low /15 & Medium /17 & High /17 &(from Equation~\ref{eq:chebyshev})\\
\midrule
AMP-4 & 44 (89.8\%) & 13 & 15 & 16 & 0.1862\\
BS-LLM & 36 (73.5\%) & 11 & 12 & 13 & 0.1959\\
AMP-2 & 36 (73.5\%) & 11 & 12 & 13 & 0.1965\\
AMP-3 & 36 (73.5\%) & 10 & 14 & 12 & 0.1987\\
EZR & 34 (69.4\%) & 10 & 13 & 11 & 0.1953\\
HKMA-scout & 33 (67.3\%) & 11 & 10 & 12 & 0.1956\\
HKMA-combined & 33 (67.3\%) & 11 & 11 & 11 & 0.1963\\
HKMA-rag & 33 (67.3\%) & 12 & 12 & 9 & 0.1971\\
Random & 23 (46.9\%) & 8 & 9 & 6 & 0.2226\\
RRP & 13 (26.5\%) & 3 & 4 & 6 & 0.2675\\
\end{tabular}
\end{widetable}

To understand why AMP-4 does so well, it is insightful to compare it against its simpler variants:
\begin{itemize}
\item
 Prompting the model to analyze feature trade-offs prior to generation (AMP-2) yields only 36 best-group memberships and a mean score of $0.1965$, matching BS-LLM (36 memberships, mean $0.1959$).
 \item
 Adding explicit constraint discovery without validation (AMP-3) also does no better than  36 memberships and slightly increases the mean Chebyshev distance to $0.1987$.
 \end{itemize}
 Hence we say that  asking an LLM to formulate domain constraints in text provides no aggregate benefit if the model is still expected to satisfy those constraints in a single generation pass. Only when each generated candidate is explicitly checked and revised against the discovered constraints (AMP-4) does performance jump from 36 to 44 best-group memberships and drop in mean Chebyshev distance by nearly a full percentage point ($0.1959$ to $0.1862$).

As to other methods: supplying external context without candidate verification failed to improve upon direct few-shot prompting. 
\begin{itemize}
\item
Combining TPE statistical scouting with document retrieval (HKMA-combined, 33 memberships, mean $0.1963$) provides no additive gain over either source alone. 
\item
For HKMA-scout and HKMA-combined, allocating 10 of the 24 evaluations to preliminary TPE scouting reduces the downstream sequential search budget from 16 steps to 6 steps, offsetting the benefit of the scouting summary.
\end{itemize}

Finally, the model-free baselines reveal the importance of benchmarking LLM warm starts against capable active learners rather than simple random sampling: while all LLM methods substantially outperform Random search (23/49, mean $0.2226$) and RRP (13/49, mean $0.2675$), the lightweight EZR active learner enters the best group on 34 datasets and achieves a lower mean Chebyshev score ($0.1953$) than BS-LLM, AMP-2, AMP-3, and all three HKMA variants. Among the compared methods, only AMP-4 outperforms both BS-LLM and EZR on best-group frequency as well as mean Chebyshev distance. 

Hence, on the question of whether domain knowledge helps an LLM propose better initial configurations, we say:
\begin{rqans}
\textbf{Only if that knowledge is checked.} Unchecked knowledge
(LLM-discovered rules, scouting, documents) did no better than direct
few-shot prompting or the EZR active learner. But AMP-4, which checks each
candidate against the rules, reached the best statistical rank on 44 of 49
datasets.
\end{rqans}

\subsection{RQ2: Performance across dimensionality}\label{sec:rq2}
\noindent{\bf RQ2:} {\bf Does increasing the number of configuration parameters change which kinds of knowledge help?}

If this were true, then any optimization oracle
(LLM or symbolic, or human) might perform worse as the number of configuration interactions increased.

To check this, we look at the {\em \#configuration options} columns of Table~\ref{tab:primary-low-label}. Partitioning the 49 datasets across low ($p < 6$), medium ($6 \le p \le 11$), and high ($p > 11$) parameter dimensionality reveals distinct scaling behaviors across knowledge strategies: some algorithms improve as dimensionality increases, while others degrade.
Algorithms that improve as the number of configuration options increases are:

\begin{itemize}
\item
AMP-4 leads every tier and exhibits its largest advantage in high-dimensional spaces, entering the best group on 13 of 15 low-dimensional tasks (86.7\%), 15 of 17 medium-dimensional tasks (88.2\%), and 16 of 17 high-dimensional tasks (94.1\%).
By comparison, direct few-shot prompting (BS-LLM) and unvalidated analysis (AMP-2) remain between 70.6\% and 76.5\% across all three tiers.
\item
HKMA-scout performs best in high-dimensional spaces relative to its other tiers, securing 12 of 17 best-group memberships (70.6\%) compared with 9 of 17 for HKMA-rag, perhaps because TPE scouting supplies concrete empirical configurations observed directly in the high-dimensional space.
\end{itemize}
On the other hand, other algorithms are clearly challenged by an increasing number of options:
\begin{itemize}
\item
Retrieval-only prompting (HKMA-rag) performs well on low-dimensional tasks (12/15, 80.0\%) and medium-dimensional tasks (12/17, 70.6\%), but degrades sharply to 9 of 17 tasks (52.9\%) in high-dimensional spaces. 
\item
Model-free methods also show sensitivity to dimensionality: Random search falls from 53.3\% (8/15) in low dimensions to 35.3\% (6/17) in high dimensions.
\item
EZR declines from 76.5\% (13/17) in medium dimensions to 64.7\% (11/17) in high dimensions.
\end{itemize}
In any case, if we focus on the methods that did best in RQ1, on the question of whether more configuration parameters change which kinds of knowledge help, we say:
\begin{rqans}
\textbf{Yes.} Checking matters most in high dimensions, where AMP-4 ranked
best on 16 of 17 tasks, while documentation-only prompting fell from 12 of
15 (low) to 9 of 17 (high).
\end{rqans}

\subsection{RQ3: Expert-reviewed knowledge}\label{sec:rq3}
\noindent{\bf RQ3:} {\bf Does human review of the LLM's domain rules help?}

Our pre-experimental intuition was that combining sources of intelligence (human and LLM) should improve our ability to configure software. If that were true, then our human-in-the-loop method (H-DKP) should outperform unguided prompting and match or exceed the automated baselines.

Table~\ref{tab:expert-low-label} compares H-DKP against the ten automated 24-label baselines across the eight expert-study datasets (nine independent expert tracks). In the eleven-method Scott--Knott analysis, H-DKP and model-free EZR tie for the highest best-group membership, each entering Group 1 on seven of the eight datasets (87.5\%), ahead of BS-LLM (6/8) and AMP-4 and AMP-3 (5/8 each). These aggregate group counts are identical whether POM3a is evaluated using Expert Track 1 or Expert Track 2.

\begin{widetable}[t]
\caption{Scott--Knott Group 1 (best-group) membership counts across the eight expert-study datasets when comparing H-DKP against the ten automated 24-label methods. Separate analyses using POM3a Expert 1 and POM3a Expert 2 yield identical best-group counts.}\label{tab:expert-low-label}
\centering\small
\setlength{\tabcolsep}{6pt}
\begin{tabular}{lrr}
Method & Best /8 & Share (\%)\\
\midrule
H-DKP & 7/8 & 87.5\%\\
EZR & 7/8 & 87.5\%\\
BS-LLM & 6/8 & 75.0\%\\
AMP-4 & 5/8 & 62.5\%\\
AMP-3 & 5/8 & 62.5\%\\
HKMA-rag & 4/8 & 50.0\%\\
HKMA-combined & 4/8 & 50.0\%\\
HKMA-scout & 3/8 & 37.5\%\\
AMP-2 & 3/8 & 37.5\%\\
Random & 3/8 & 37.5\%\\
RRP & 2/8 & 25.0\%
\end{tabular}
\end{widetable}

Because eight datasets with 20 trials each can place multiple methods into the same Scott--Knott cluster when within-task variance is high, we also inspect trial medians on the individual tracks to separate central tendency from outlier trials. Even where multiple methods share Group 1 on mean score, H-DKP achieves a lower 20-trial median Chebyshev distance than both BS-LLM and EZR on all seven non-tied datasets (and ties at the $0.2500$ optimum on Auto93). Inspecting individual tasks also explains why AMP-4 enters Group 1 on five of the eight expert datasets compared with seven for H-DKP: on both Life Expectancy ($p=20$) and LLVM ($p=10$), AMP-4 matches or beats BS-LLM and EZR in median shortfall ($0.0821$ on Life Expectancy, tying H-DKP versus $0.1014$ for BS-LLM and $0.0966$ for EZR; $0.0151$ on LLVM versus $0.0163$ for BS-LLM and $0.0194$ for EZR), but occasional high-shortfall outlier trials raise its mean score into Scott--Knott Group 2. Conditioning every run on a frozen set of expert-verified rules in H-DKP suppresses these outlier runs while preserving low median shortfall.

Examining the expert elicitation logs shows why this curation works. Across the nine tracks, specialists audited 61 initial zero-shot LLM claims, classifying 47 as \textsc{valid} (77.0\%), 8 as \textsc{modify} (13.1\%), 1 as \textsc{invalid} (1.6\%), and 5 as \textsc{unsure} (8.2\%), which together with confirmed follow-up additions yielded 56 frozen rules (2 to 8 rules per track). Rather than rubber-stamping model beliefs, experts actively filtered flawed assumptions (Box~1), revealing two practical patterns:
\begin{itemize}
\item \emph{Rule precision outweighs quantity:} On XOMO Flight ($p=23$, the highest-dimensional task in the cohort), the specialist confirmed only two rules and marked two others uncertain, yet those two constraints guided H-DKP to a median score of $0.1138$, beating BS-LLM ($0.1239$), EZR ($0.1231$), and AMP-4 ($0.1172$).
\item \emph{Equifinality across independent specialists:} On POM3a ($p=9$), two experts reviewed the same initial LLM claims and produced different rule sets (Expert 1 validated six and modified two for eight rules; Expert 2 validated six and marked two uncertain for six rules). Despite generating different initial warm starts, both tracks converged to the exact same median benchmark configuration ($D = 0.1239$), improving substantially over BS-LLM ($0.1648$) and EZR ($0.1547$).
\end{itemize}

Hence, on the question of whether human review of the LLM's domain rules helps, we say:
\begin{rqans}
\textbf{Yes.} H-DKP ranked best on 7 of 8 expert tasks, and beat direct
prompting on 7 of 8. However, EZR also ranked best on 7 of 8, at zero cost.
\end{rqans}

\subsection{RQ4: Cost and budget-matched sensitivity}\label{sec:rq4}
\noindent{\bf RQ4:} {\bf What does checked knowledge cost?}

A pressing concern with LLM-based methods is the evaluation and computational cost associated with running them. In particular, prior work such as SynthCore \citep{senthilkumar2026synthcore} pools 16 initial representatives with 15 few-shot generation sessions (76 nominal requests, averaging 55 distinct labels per trial), consuming more than twice the evaluation budget of our 24-label methods.

To test whether multi-session ensembling remains competitive without an expanded labeling budget, Table~\ref{tab:synthcore28-sensitivity} compares the budget-matched three-session restriction, SynthCore-28 (28 nominal requests; 26.31 mean distinct labels, range 20--32), against the ten 24-label methods across all 49 datasets. When restricted to roughly the same evaluation budget as the 24-label methods, SynthCore-28 drops to ninth place, entering the best Scott--Knott group on only 23 of 49 datasets with a mean Chebyshev distance of $0.2063$. Under equalized budgets, AMP-4 (39/49 in this eleven-method partitioning, mean $0.1862$), BS-LLM (34/49, mean $0.1959$), and model-free EZR (32/49, mean $0.1953$) all substantially outperform SynthCore-28. Spending scarce labels on 16 initial representatives and unguided few-shot sessions without sequential search is therefore far less sample-efficient than pairing a single validated warm start with sequential optimization.

\begin{widetable}[t]
\caption{Budget-matched sensitivity analysis comparing three-session SynthCore-28 (28 nominal requests, 26.31 mean distinct labels) against the ten 24-label methods across 49 datasets. Scott--Knott groups are recomputed across all eleven methods.}\label{tab:synthcore28-sensitivity}
\centering\small
\setlength{\tabcolsep}{5pt}
\begin{tabular}{lrrr}
Method & Distinct labels & Best /49 & Mean $D$\\
\midrule
AMP-4 & $\le 24$ & 39 & 0.1862\\
BS-LLM & $\le 24$ & 34 & 0.1959\\
EZR & 24 & 32 & 0.1953\\
AMP-2 & $\le 24$ & 32 & 0.1965\\
AMP-3 & $\le 24$ & 32 & 0.1987\\
HKMA-rag & $\le 24$ & 29 & 0.1971\\
HKMA-scout & $\le 24$ & 28 & 0.1956\\
HKMA-combined & $\le 24$ & 26 & 0.1963\\
SynthCore-28 & 26.31 & 23 & 0.2063\\
Random & 24 & 19 & 0.2226\\
RRP & Dataset-dep. & 12 & 0.2675
\end{tabular}
\end{widetable}

Beyond configuration evaluations, the methods differ in their computational and human preparation overheads (Table~\ref{tab:repeat-cost-revised}). Model-free baselines (Random, EZR, and RRP) incur zero LLM token costs and complete in a quarter of a second or less ($\le 0.23$~s). Among LLM-assisted methods, single-call approaches incur the lowest API charges per trial: HKMA-rag averages 1,219 tokens (0.02\textcent) and BS-LLM averages 1,688 tokens (0.03\textcent). Staged prompting increases token consumption alongside the number of reasoning steps: AMP-2 averages 4,648 tokens (0.07\textcent\ across 2 calls), AMP-3 averages 10,279 tokens (0.15\textcent\ across 3 calls), and AMP-4 averages 25,567 tokens (0.54\textcent\ across 7 calls).

\begin{widetable}[t]
\caption{LLM calls and mean computational and inference cost per completed trial across the ten 24-label methods with GPT-OSS-120B, ordered by monetary API charge in US cents (\textcent). Tokens sum input and output tokens.}\label{tab:repeat-cost-revised}
\centering\small
\setlength{\tabcolsep}{5pt}
\begin{tabular}{lrrrr}
Method & LLM calls & Tokens & API cost (\textcent) & Time (s)\\
\midrule
Random & 0 & 0 & 0 & $<$0.01\\
EZR & 0 & 0 & 0 & 0.04\\
RRP & 0 & 0 & 0 & 0.23\\
HKMA-rag & 1 & 1,219 & 0.02 & 9.20\\
BS-LLM & 1 & 1,688 & 0.03 & 2.40\\
HKMA-scout & 2 & 4,033 & 0.04 & 21.10\\
HKMA-combined & 2 & 4,171 & 0.05 & 20.20\\
AMP-2 & 2 & 4,648 & 0.07 & 39.80\\
AMP-3 & 3 & 10,279 & 0.15 & 80.30\\
AMP-4 & 7 & 25,567 & 0.54 & 209.80
\end{tabular}
\end{widetable}

Although AMP-4 requires the largest token budget among the tested methods, its absolute inference cost remains roughly half of one US cent (0.54\textcent) and 3.5 minutes (209.80~s) of sequential API latency per trial. For H-DKP, upfront offline preparation required 6--10 short email exchanges (1--3 lines per follow-up reply over 5--17 calendar days) and an estimated 6--48 nominal benchmark accesses per track to formulate diagnostic questions (Section~\ref{sec:human-study}); once frozen, the rule set is reusable across optimization runs, requiring only a single online LLM call per trial comparable in token cost to BS-LLM.

Hence, on the question of what checked knowledge costs, we say:
\begin{rqans}
\textbf{Very little.} AMP-4 costs half a cent of LLM calls per trial ($\le 24$ labels) and H-DKP needs only 6 to 10 short expert emails. By contrast, restricting multi-session SynthCore to a comparable budget (SynthCore-28, 26.31 labels) drops its win rate to 23 of 49 tasks, trailing both AMP-4 (39 of 49) and zero-cost EZR (32 of 49).
\end{rqans}

\section{Threats to validity}\label{sec:threats}

\subsection{Construct validity}
We evaluate final optimization quality using Chebyshev ($L_\infty$) distance to the ideal point, which penalizes the worst normalized shortfall across all objectives. Different multi-objective indicators, such as hypervolume or Euclidean distance to heaven, reflect alternative practitioner preferences over Pareto trade-offs \citep{Li22}. Moreover, the downstream sequential optimizer (SMO), EZR, and SynthCore-28 rank candidates internally using Euclidean distance to heaven. Scoring the final configuration returned by every method with Chebyshev distance provides a uniform external criterion that requires balanced attainment across all objectives, although aligning internal search objectives with the external evaluation metric could further improve final scores.

Because MOOT is a collection of finite tabular benchmarks, objective normalization uses the minimum and maximum values recorded in each reference table, and LLM-generated configurations are evaluated by mapping them to the nearest observed row in normalized feature space. When a proposed configuration lies far from recorded table entries, nearest-neighbor lookup is only an approximation of true system behavior \citep{pmlr-v188-pfisterer22a,zela2022surrogatenasbenchmarksgoing}. To prevent tabular approximation artifacts from misleading human specialists during the H-DKP study, we inspected proposal-to-row distances prior to each contact and switched from failure-diagnosis queries to profile-plausibility queries whenever the nearest observed row was too distant to represent the generated candidate faithfully.

\subsection{Internal validity}
The compared methods differ in their prompt structures, knowledge sources, and labeling budgets, all of which contribute to final optimization performance. To prevent sample size from confounding the comparison of knowledge strategies, our primary benchmark evaluation equalizes the total budget at 24 configuration evaluations across Random search, EZR, BS-LLM, the three AMP variants, the three HKMA variants, and H-DKP (with RRP governed by its dataset-dependent recursive partitioning depth). For the budget-matched SynthCore-28 baseline, we explicitly track and report both total evaluation requests and mean distinct configurations labeled per trial. In H-DKP, the offline rule-elicitation phase used 36 generation-associated exchanges (an estimated 6--48 nominal table accesses per track, or 216 in total; Table~\ref{tab:expert-cohort}) to construct diagnostic questions before freezing the rules; we report these one-time offline preparation costs separately from the 24-label online trial budget, and note that our effort logs record email counts, reply line lengths, and calendar turnaround (5--17 days) rather than active working minutes. Moreover, downstream H-DKP performance does not depend on heavy offline table querying: on XOMO Flight ($p=23$) and POM3a Track 1 ($p=9$), Phase A required only one and two generation-associated exchanges (6 and 12 nominal table accesses, respectively; Table~\ref{tab:expert-cohort}), yet both tracks substantially outperformed BS-LLM and EZR across the 20 online trials.

Using pre-trained LLMs introduces potential threats from benchmark memorization, provider updates, and stochastic decoding variance \citep{sallou2024breaking}. Although GPT-OSS-120B may have been exposed to public documentation or descriptions of standard software systems during pre-training, all LLM baselines and knowledge-guided variants use the exact same underlying model and decoding parameters, ensuring that relative differences reflect the prompting and knowledge workflows rather than model selection. To account for stochastic generation and occasional malformed model outputs, we execute 20 independent trials per dataset and log all prompts, responses, and random seeds. When an LLM call returns unmappable feature values or fails schema parsing, the evaluation pipelines apply deterministic fallback rules (for example, SynthCore-28 falls back to a session's initial representatives if no valid candidates are parsed in that session, and the AMP-4 evaluator retains prior valid settings when a revised field cannot be mapped). Consequently, the reported results reflect the end-to-end performance of each prompting and validation pipeline across completed trials.

\subsection{Conclusion validity}
Within each dataset, we group methods using Scott--Knott clustering gated by a non-parametric effect size threshold (Cliff's $|\delta| \ge 0.147$) and a studentized bootstrap test ($p < 0.05$). All automated and expert-study comparisons are based on 20 trials per dataset. Because Scott--Knott partitions depend on the specific pool of methods included in a comparison, we recompute clusters separately for the ten-method primary scope, the eleven-method expert scope, and the SynthCore-28 sensitivity scope. Sharing Group 1 on a dataset indicates that differences among those methods are either statistically insignificant or small in effect size; we therefore report equal-weighted mean Chebyshev distances and trial medians alongside Group 1 membership counts to capture both statistical rank and shortfall magnitude.

\subsection{External validity}
Our 49-dataset benchmark corpus combines measured software system configurations (such as compilers, databases, and machine learning pipelines), software process simulations, and general optimization tasks from the MOOT repository. While this diverse corpus enables a broad evaluation across low, medium, and high parameter dimensionalities, dimensionality covaries with problem domain, table cardinality, and feature types across the benchmark. Furthermore, evaluating configurations via tabular lookup does not capture live execution variance, hardware interference, or workload drift on deployed production systems. Finally, recruitment for the H-DKP study depended on access to specialists with domain expertise in the target tasks. While our nine expert tracks across eight datasets (spanning 2 to 30 years of experience, including both system-specific authors/maintainers and general software engineering or quantitative researchers; Table~\ref{tab:expert-cohort}) provide initial empirical evidence on human-in-the-loop rule curation, broader studies with larger pools of practitioners are needed to characterize inter-expert variability across industrial software systems.

\section{Discussion}\label{sec:discussion}

Our main result is simple: checking an LLM's domain knowledge matters more than adding to it. Rules, documents, and scouted data did not help until each candidate was checked against them, either by the LLM itself (AMP-4) or by a human expert (H-DKP). The rest of this section explains why, what experts add, and what this means for practitioners.

\subsection{Why constraint verification outperforms passive context injection}
A central finding of our study is that domain knowledge improves LLM warm starts only when that knowledge is actively checked, either by the model itself (AMP-4) or by a human specialist (H-DKP). Under a fixed 24-label budget, prompting the model to analyze feature trade-offs (AMP-2, 36/49 best-group memberships, mean $0.1965$) or to explicitly list parameter constraints before generation (AMP-3, 36/49, mean $0.1987$) yields no aggregate improvement over direct few-shot prompting (BS-LLM, 36/49, mean $0.1959$). However, when AMP-4 validates each generated candidate individually against those same model-discovered constraints and revises violations, best-group membership rises to 44 of 49 datasets (89.8\%) and mean Chebyshev distance drops to $0.1862$.

This contrast between AMP-3 and AMP-4 highlights an instruction-following gap between natural-language reasoning and numerical vector generation. An LLM can articulate sensible parameter dependencies in text, yet still violate its own stated constraints when asked to output four multi-dimensional configurations simultaneously in a single pass. Separating generation from per-candidate verification allows the model to audit one configuration vector at a time and repair infeasible or conflicting parameter assignments before physical evaluations are spent. This mechanism also explains why AMP-4 scales effectively with dimensionality, winning on 16 of 17 high-dimensional datasets ($p > 11$, 94.1\%) whereas unvalidated constraint discovery (AMP-3) drops from 14 of 17 in medium dimensions to 12 of 17 in high dimensions.

The HKMA results reinforce the limitation of passive context injection. All three HKMA variants enter the best group on 33 of 49 datasets, trailing BS-LLM (36/49) despite incorporating retrieved documentation, statistical scouting summaries, or both. Two opposing trade-offs explain this outcome:
\begin{enumerate}
    \item \emph{Scouting versus sequential search allocation:} Under a strict 24-label cap, HKMA-scout and HKMA-combined spend 10 evaluations on TPE scouting prior to LLM invocation, leaving only 6 evaluations for downstream sequential optimization (compared with 16 SMO steps for BS-LLM and AMP). While scouting helps the LLM observe empirical patterns in high-dimensional spaces (where HKMA-scout achieves 12/17 best-group memberships versus 9/17 for HKMA-rag), starving the sequential optimizer of post-warm-start iterations offsets this advantage across the broader benchmark.
    \item \emph{Documentation specificity in high dimensions:} Retrieval-only prompting (HKMA-rag) preserves all 16 SMO steps and performs well on low-dimensional tasks (12/15, 80.0\%), but falls to 9 of 17 (52.9\%) on high-dimensional tasks. General documentation passages rarely specify joint settings across more than a dozen interacting parameters, and combining scouting summaries with retrieved text (HKMA-combined) adds prompt length without enforcing constraint satisfaction.
\end{enumerate}

\subsection{Human expertise, rule curation, and model-free baselines}
The H-DKP study shows why active curation matters when incorporating human domain knowledge. Across the nine expert tracks, specialists did not merely rubber-stamp the LLM's zero-temperature beliefs; they modified claims on five tracks, rejected an invalid claim on PostgreSQL, and flagged uncertain claims on three tracks. Incorporating these verified rules enabled H-DKP to achieve a lower median Chebyshev distance than direct prompting (BS-LLM) on seven of the eight tasks (tying on Auto93) and to enter the top Scott--Knott group on seven of eight tasks. Moreover, comparing H-DKP with AMP-3 and AMP-4 clarifies how human rule curation and automated candidate verification address the instruction-following gap in complementary ways. In AMP-3, the LLM formulates unconstrained rule lists and attempts to satisfy them in a single pass, often violating its own stated dependencies. Although H-DKP also uses single-pass generation during online trials (Phase B), it avoids AMP-3's degradation for two reasons: (1)~during offline elicitation (Phase A), candidate rules are iteratively tested against intermediate proposals and failure-diagnosis queries (Algorithm~\ref{alg:hdkp}, lines 5--9), filtering out rules that the model misapplies before the knowledge base $K$ is frozen; and (2)~specialists prune the knowledge base down to a compact set of two to eight relevant constraints per track, reducing instruction-following overload during generation and stabilizing trial-to-trial variance relative to self-discovered rules in AMP-4.

The expert tracks also demonstrate that the precision of domain constraints matters more than their quantity, and that distinct expert perspectives can converge to equivalent optimization outcomes:
\begin{itemize}
    \item \emph{High-leverage constraints in large spaces:} On XOMO Flight ($p=23$), the specialist confirmed only two initial rules and marked two others uncertain. Despite leaving most of the 23 parameters unconstrained by explicit rules, those two verified constraints guided H-DKP to the lowest median Chebyshev score ($0.1138$) among all compared methods.
    \item \emph{Equifinality across independent experts:} On POM3a ($p=9$), two specialists independently reviewed the same initial LLM claims and produced distinct rule sets: Expert 1 confirmed eight rules (including two modified conditions), whereas Expert 2 confirmed six rules and marked two uncertain. Although these two knowledge bases produced different initial warm-start proposals and distinct trial distributions, subsequent sequential search guided both tracks to the exact same median benchmark configuration ($D = 0.1239$), outperforming both BS-LLM ($0.1648$) and EZR ($0.1547$). Different experts can therefore emphasize different valid domain relationships while steering sequential optimization into the same Pareto-optimal basin.
\end{itemize}

At the same time, evaluating LLM warm starts alongside the model-free EZR active learner provides an essential reality check. Many studies of LLMs for optimization compare primarily against Random search. Had we compared only against Random search (23/49 best-group memberships, mean $0.2226$), every LLM prompting variant would appear to offer a substantial gain. In contrast, EZR uses the same 24-label budget at zero token cost and 0.04 seconds per trial, yet achieves a lower mean Chebyshev score ($0.1953$) than BS-LLM ($0.1959$), AMP-2 ($0.1965$), AMP-3 ($0.1987$), and all three HKMA variants, while matching H-DKP's best-group count (7/8) on the expert cohort. Consequently, unverified LLM prompting barely improves upon a well-designed active-learning heuristic. Only active constraint validation (AMP-4 across the 49 datasets) and human rule curation (H-DKP on trial medians) consistently justify the added complexity of invoking a language model.

\subsection{Trading inference compute against physical system evaluations}
Comparing AMP-4 ($\le 24$ labels, 0.54\textcent\ per trial) against budget-matched SynthCore-28 (26.31 mean distinct labels) and model-free EZR (24 labels, 0\textcent\ per trial) clarifies the trade-off between LLM inference compute and physical configuration evaluations. While original SynthCore \citep{senthilkumar2026synthcore} relies on pooling 16 initial representatives with 15 generation sessions (averaging 55 distinct configuration labels per trial), our budget-matched sensitivity analysis (Table~\ref{tab:synthcore28-sensitivity}) shows that multi-session ensembling breaks down when restricted to roughly two dozen evaluations. Using 26.31 mean distinct labels, SynthCore-28 enters the best statistical group on only 23 of 49 datasets (mean $D = 0.2063$), losing to 24-label AMP-4 (39/49, mean $0.1862$), BS-LLM (34/49, mean $0.1959$), and zero-cost EZR (32/49, mean $0.1953$). Without sequential search feedback or constraint validation, allocating 16 labels to initial partitioning and 12 labels to unguided few-shot proposals uses a small evaluation budget inefficiently.

At the same time, the inference cost profile in Table~\ref{tab:repeat-cost-revised} places AMP-4's computational overhead relative to EZR. Whereas EZR requires zero API tokens and completes in 0.04 seconds, AMP-4 requires 7 LLM calls and 25,567 tokens per trial, corresponding to an average API cost of 0.54\textcent\ (roughly half of one US cent) and 3.5 minutes of sequential inference time. When LLM access is unavailable or sub-second latency is required, EZR provides a strong model-free default that matches or beats every unverified prompting strategy. However, in real-world software engineering tasks, such as compiling a large codebase like LLVM, executing database benchmarks in PostgreSQL, or training machine learning models, evaluating a single candidate configuration on the target system can require several minutes to several hours of compute time. Spending half a cent and 3.5 minutes of LLM inference on staged constraint validation to secure top-ranked configurations within 24 labels (39/49 in the sensitivity scope and 44/49 in the primary scope, versus 32/49 and 34/49 for EZR) represents a favorable trade-off whenever physical system evaluations dominate total runtime.

\section{Conclusion}\label{sec:conclusion}

Configuring complex software is an important and difficult task: parameter spaces are large, configuration errors are common, and each evaluation can be slow and costly. Prior work on LLM warm starts, such as SynthCore~\citep{senthilkumar2026synthcore}, had two drawbacks. It was fully automatic, with no human in the loop, and it needed dozens of distinct labels per trial. Here, we studied what an LLM should know before it proposes its first configurations, using just 24 labels, and we tested human experts as one source of that knowledge. 

Across 49 datasets and nine expert tracks, our lead result is that domain knowledge helps only when it is checked. AMP-4, where the LLM checks each candidate against its own rules, reached the best statistical rank on 44 of 49 datasets. H-DKP, where human experts correct the LLM's rules, did so on 7 of 8 expert tasks.
 
Our empirical findings translate into concrete guidelines for selecting an optimization strategy based on evaluation cost, parameter dimensionality, and available domain assets. When system evaluations are expensive and labels are strictly capped ($\le 24$), use AMP-4. It requires no external documentation or human expert time, costs roughly half a cent per trial in LLM inference, and enters the top statistical group on 89.8\% of tasks overall and 94.1\% of high-dimensional tasks. When LLM access or inference latency is restricted, use model-free EZR instead. It executes in milliseconds, requires no API tokens, and matches or outperforms unvalidated LLM prompting (BS-LLM, AMP-2, AMP-3, and HKMA) in mean optimization quality. Finally, when domain specialists are available for recurring tuning tasks, use H-DKP. Although asynchronous elicitation requires an upfront investment of six to ten expert contacts, the resulting frozen rule set requires only a single LLM call per run, can be reused across repeated optimization campaigns, and consistently lowers median shortfall relative to unguided prompting.

This study suggests several directions for future work. First, our human-in-the-loop study covers eight datasets and nine expert tracks; larger expert cohorts would test whether the H-DKP results hold across more domains and more specialists per task. Second, all LLM-guided methods here use GPT-OSS-120B. Repeating the study with other models would show whether the gap between checked and unchecked knowledge is a property of the method or of one model. Third, AMP-4 checks only the four initial candidates, but the same per-candidate validation could also guide the sequential optimizer that selects the remaining labels. Lastly, MOOT tasks are tabular benchmarks, where each label is a table lookup. Future work should apply these methods to live systems, where each label is a real compile, benchmark run, or training job.

\backmatter
\section*{Statements and declarations}
\subsection*{Data and code availability}
The study uses tabular tasks from MOOT \citep{mootrepo}. Available at: 
\url{https://github.com/timm/moot}

The replication package is available at: \url{https://github.com/srina1h/What-Should-an-LLM-Know}

\begingroup
\raggedright
\bibliography{references}
\endgroup
\end{document}